\documentclass[ aip,jcp,reprint,groupedaddress, amssymb, amsmath]{revtex4-1}
\usepackage{amssymb, amsmath,amsfonts}
\usepackage{graphicx}% Include figure files
\usepackage{dcolumn}% Align table columns on decimal point
\usepackage{bm}% bold math
\usepackage{multirow}
\usepackage{mathtools}

\begin{document}
\newcommand{\rmsub}[2]{\ensuremath{#1_{\rm #2}}}
\newcommand{\rmsubsub}[3]{\ensuremath{{#1_{\rm #2}}_{#3}}}

\title{Coarse-grained Modeling of DNA Curvature}

\author{Gordon S. Freeman}
\affiliation{Department of Chemical and Biological Engineering\\ University of Wisconsin--Madison, Madison, WI, 53706 USA}
\author{Daniel M. Hinckley}
\affiliation{Department of Chemical and Biological Engineering\\ University of Wisconsin--Madison, Madison, WI, 53706 USA}

%\affiliation{Department of Chemical and Biological Engineering\\ University of Wisconsin--Madison, Madison, WI, USA}
\author{Joshua P. Lequieu}
\affiliation{Institute for Molecular Engineering\\ University of Chicago, Chicago, IL 60637 USA}
\author{Jonathan K. Whitmer}
\affiliation{Materials Science Division\\ Argonne National Laboratory, Argonne, IL 60439 USA}
\author{Juan J. de Pablo}
\affiliation{Institute for Molecular Engineering\\ University of Chicago, Chicago, IL 60637 USA}
\affiliation{Materials Science Division\\ Argonne National Laboratory, Argonne, IL 60439 USA}
\email{Corresponding Author: depablo@uchicago.edu}

\date{\today}

\begin{abstract}
Modeling of DNA--protein interactions is a complex process involving many important time and length scales. 
This can be facilitated through the use of coarse-grained models which reduce the number of degrees of freedom and allow efficient exploration of binding configurations. 
It is known that the local structure of DNA can significantly affect its protein-binding properties (i.e. intrinsic curvature in DNA-histone complexes).
In a step towards comprehensive DNA--protein modeling, we expand the 3SPN.2 coarse-grained model to include intrinsic shape, and validate the refined model against experimental data including melting temperature, local flexibility, persistence length, and minor groove width profile.
\end{abstract}
\maketitle

\section{Introduction}

Coarse-grained (CG) models provide access to time and length scales that are not generally accessible to all-atom (AA) molecular simulations.
CG models have been applied to a wide variety of systems, including liquid crystals, block copolymers, proteins, and dexoxyribonucleic acid (DNA).
Here we confine our discussion to CG models of DNA, which have been used to study various phenomena, including hybridization\cite{Hinckley2014}, stretching\cite{Romano2013}, and bubble formation\cite{Zeida2012}. 
Several DNA CG models have been proposed in recent years\cite{Dans2010,Ouldridge2011,Edens2012, He2013, Hinckley2013}; 
%to the best of our knowledge, no DNA CG model has yet been applied to the study of DNA--protein interactions.
to the best of our knowledge, models designed specifically for studies of DNA structure and thermodynamic properties have not been applied to explore DNA--protein interactions.

Proteins do not have direct access to the functional groups that lead to the formation of the Watson--Crick (W--C) base pairs;
however, the sequence of contiguous W--C base pairs results in local deviations from the mean double helix that facilitate the binding of a protein to a unique location along the double-stranded DNA (dsDNA).
These changes can be quantified by the intrinsic flexibility and geometry of sequential nucleotide pairs (a base-step\cite{Olson1998}) or quartets\cite{Lavery2010}.
Variations in the widths of the major and minor grooves, as well as the flexibility of individual base-steps, dictate the energetic benefit of binding.
A CG model suitable for modeling DNA interacting with proteins should capture these sequence effects.
In addition, the CG model should include explicit electrostatics, as charged amino acid side chains interact with the negatively--charged DNA backbone.
Most CG models include sequence--dependence to some degree, usually in the CG topology and the energy parameters of the base stacking and base pairing interactions.
Some models have demonstrated consistency with local base-step properties such as rise and twist\cite{Dans2010, Edens2012}.
However, no CG model to date has been demonstrated to capture the intrinsic curvature and global flexibility of DNA.

We present 3SPN.2C, a modification of the 3SPN.2 CG DNA model\cite{Hinckley2013}.
3SPN.2 was previously shown to correctly model DNA biophysics by penalizing deviations from the ideal B-form DNA (B--DNA) crystallographic structure\cite{Arnott1976}.
The use of ideal B--DNA to define the minimum energy configuration suggests that sequence effects can be included by using sequence--specific base-step parameters to build the configuration.
Additional sequence--dependence can be added by making the flexibility of bonded interactions dependent on the sequence context.
Here we adopt both of these changes to extend the 3SPN.2 model to include sequence-dependent shape and flexibility.

The manuscript begins with a description of the model and the data and methods used to assign sequence-dependent parameters.
Results are then presented to demonstrate consistency with experimental melting temperatures and flexibilities.
Lastly, we present a comparison of simulated minor groove widths to available experimental data.

\section{Methods}

\subsection{3SPN.2C DNA Model}

The 3SPN.2C model represents an extension of 3SPN.2, a third-generation CG model\cite{Knotts2007,Sambriski2009a,Hinckley2013} shown recently to capture the correct structural, mechanical, and thermodynamic properties of DNA.
The bonded and non-bonded potentials of 3SPN.2 were constructed to penalize deviations from the B-DNA crystal structure\cite{Arnott1976}.
Force constants of the bond, angle, and dihedral potentials were independent of sequence, while base stacking energies and base pairing energies were sequence--dependent.
Special emphasis was placed on capturing the correct flexibilities of both single--stranded DNA (ssDNA) and dsDNA.
3SPN.2C is intended for simulations of dsDNA interacting with proteins.
Consequently, emphasis is placed on capturing the sequence-dependent shape and flexibility of dsDNA.
The properties of ssDNA are not prioritized, making 3SPN.2 better suited for studies involving ssDNA. 

In the 3SPN.2C model, the original version of 3SPN.2 is modified as follows:
First, the reference configuration that defines the minimum energy structure of dsDNA is modified to include sequence-dependent shape.  
The methodology for so doing is described in Section \ref{BaseStepParameters}.
This results in equilibrium distances and angles that are a function of the base-step\footnote{Code for generating initial topologies and simulating 3SPN.2C is available upon request}.
Second, each base step is assigned unique force constants for each bend angle, as described in Section \ref{BaseStepFlexibility}.  
Lastly, weak dihedral potentials are assigned to all dihedral angles formed by the three-site-per-nucleotide topology.
These dihedrals, with the functional form
\begin{equation}\label{eq:periodic}
U_{\phi,\mbox{\tiny{periodic}}} = k_{\phi,\mbox{\tiny{periodic}}} \left[ 1+ \text{cos}\left(\phi - \phi_0\right)\right],
\end{equation}
provide additional stability to the helix when deformed severely from the equilibrium structure, as is often the case when DNA binds to protein (i.e. DNA-histone binding).
The magnitude of the torsion force constants, which are independent of sequence, is modified to provide qualitative agreement with experimental data from Ref. \onlinecite{Geggier2010}.

The aforementioned modifications require the bonded and non-bonded energy parameters to be adjusted in order to preserve agreement with experimental data.
Metadynamics simulations are used to assign both the intrastrand stacking energies and interstrand base pairing and cross stacking energies, as in the original parameterization of 3SPN.2\cite{Hinckley2013}.
The force constants and energy parameters for all interactions can be found in the Appendix.
{All simulations but the melting temperature calculations (cf. Fig. \ref{Sec:Tm}) are performed at 300 K and 150 mM ionic strength in the NVT ensemble using a Langevin thermostat.

\subsection{Imparting Sequence-dependent Shape} \label{BaseStepParameters}

\begin{figure*}
  \begin{center}
    \includegraphics[width=17cm]{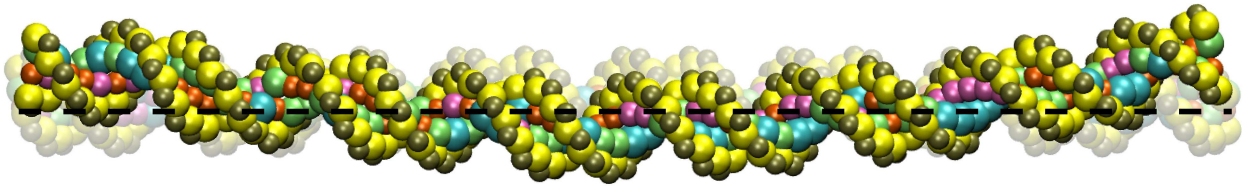}
    \caption{%
    Comparison of aligned minimum energy topologies used in 3SPN.2C (opaque) and 3SPN.2 (transparent).
    The dotted lines represents the helical axis of the topology without sequence-dependence.
    Sequence: 5'-CTGGAGAATCCCGGTGCCGAGGCCGCTCAATGGATCCTTGCAAGCTCTTGGTGCGCTTTTTCGGCTGTTGACGC-3'.
    This sequence was taken from a larger sequence shown to have high curvature (Sequence d1 in Ref. \onlinecite{Segal2006}).
    }
    \label{fig:curved}
  \end{center}
\end{figure*}

Recent studies have suggested that DNA shape is an essential component of DNA--protein recognition. 
In particular, sequence attributes such as the minor groove width and the intrinsic curvature have been shown to play a role\cite{West2010,Rohs2009,Scipioni2004,Peters2010}.
Building on such studies, we incorporate sequence-dependent shape into the 3SPN.2 model.

To implement sequence-dependent DNA shape, we followed the studies of Olson and coworkers\cite{Olson1998}, who determined average base pair and base-step parameters for each of the ten unique DNA base-steps.
For this work, we employed more recent average base-step parameters that were demonstrated to be well-suited for DNA--protein binding (Table \ref{morozov_parameters})\cite{Morozov2009}. 
Base pair parameters, shown in Table \ref{OlsonParameters}, are from Ref. \onlinecite{Olson2006}.
It should be noted that employing base-step parameters (rather than longer range approaches such as trimeric parameters) has been shown to produce good agreement with experimental electrophoretic gel retardation studies\cite{DeSantis1990}.

In order to translate these base pair and base-step parameters into an actual structure for a given DNA sequence, we used X3DNA\cite{Lu2003} to build an AA structure. 
We then coarse-grained the AA structure into 3SPN.2C by mapping each base, sugar, and phosphate to a bead placed at the respective center of mass.
The result is a topology with sequence-dependent groove widths and intrinsic curvature, as shown in Fig. \ref{fig:curved}.
The lengths and angles of each bond, bend, and dihedrals in this structure represent the minimum energy values of these interactions.
Minimum energy distances and angles of base stacking, base pairing, and cross stacking interactions were also obtained from this structure.
%Electrostatic terms are not altered in any way from 3SPN.2\cite{Hinckley2013}. 

\begin{table}
  \small
  \caption{
        Average base-step parameters for X3DNA\cite{Lu2003} required to obtain the minimum energy configuration in 3SPN.2C.
        These values are from Ref. \onlinecite{Morozov2009}.
  }\label{morozov_parameters}
  \centering
  \begin{tabular}{@{\vrule height 10.5pt depth4pt width0pt}lcccccc}
    \\[1ex]
    \hline
    Base-Step & Twist & Roll & Tilt & Shift & Slide & Rise \\
     & $^{\circ}$ & $^{\circ}$ & $^{\circ}$ & \AA & \AA & \AA \\
    \hline
        AA & 35.31 & 0.76 & -1.84 & -0.05 & -0.21 & 3.27 \\
        AT & 31.21 & -1.39 & 0.00 & 0.00 & -0.56 & 3.39 \\
        AC & 31.52 & 0.91 & -0.64 & 0.21 & -0.54 & 3.39 \\
        AG & 33.05 & 3.15 & -1.48 & 0.12 & -0.27 & 3.38 \\
        TA & 36.20 & 5.25 & 0.00 & 0.00 & 0.03 & 3.34 \\
        TT & 35.31 & 0.76 & 1.84 & 0.05 & -0.21 & 3.27 \\
        TC & 34.80 & 3.87 & 1.52 & 0.27 & -0.03 & 3.35 \\
        TG & 35.02 & 5.95 & 0.05 & 0.16 & 0.18 & 3.38 \\
        CA & 35.02 & 5.95 & -0.05 & -0.16 & 0.18 & 3.38 \\
        CT & 33.05 & 3.15 & 1.48 & -0.12 & -0.27 & 3.38 \\
        CC & 33.17 & 3.86 & 0.40 & 0.02 & -0.47 & 3.28 \\
        CG & 35.30 & 4.29 & 0.00 & 0.00 & 0.57 & 3.49 \\
        GA & 34.80 & 3.87 & -1.52 & -0.27 & -0.03 & 3.35 \\
        GT & 31.52 & 0.91 & 0.64 & -0.21 & -0.54 & 3.39 \\
        GC & 34.38 & 0.67 & 0.00 & 0.00 & -0.07 & 3.38 \\
        GG & 33.17 & 3.86 & -0.40 & -0.02 & -0.47 & 3.28 \\
    \hline
  \end{tabular}
\end{table}

\begin{table}
  \small
  \caption{
        Average base pair parameters used to construct the 3SPN.2C model.
        These values are from Ref. \onlinecite{Olson2006}.
  }\label{OlsonParameters}
  \centering
  \begin{tabular}{@{\vrule height 10.5pt depth4pt width0pt}lcccccc}
    \\[1ex]
    \hline
    base pair & Buckle & Propeller & Opening & Shear & Stretch & Stagger \\
     & $^{\circ}$ & $^{\circ}$ & $^{\circ}$ & \AA & \AA & \AA \\
    \hline
        A--T & 1.8 & -15.0 & 1.5 & 0.07 & -0.19 & 0.07 \\
        T--A & -1.8 & -15.0 & 1.5 & -0.07 & -0.19 & 0.07 \\
        G--C & 4.9 & -8.7 & -0.6 & -0.16 & -0.17 & 0.15 \\
        C--G & -4.9 & -8.7 & -0.6 & 0.16 & -0.17 & 0.15 \\
    \hline
  \end{tabular}
\end{table}

\subsection{Local Base-Step Flexibility}\label{BaseStepFlexibility}

The local flexibility of DNA base-steps has been the subject of significant interest over the past several decades\cite{Olson1998,Morozov2009,Olson2006}. 
%DNA sequence has long been known to play a role in DNA--protein recognition, and many attempts have been made to elucidate the mechanism through which this ``indirect readout'' process occurs. 
Olson and coworkers mined structural databases and calculated the relative flexibilities of DNA base-steps from these data \cite{Olson1998}. 
In order to perform this analysis on the atomistic representation of DNA models, the many degrees of freedom (DOF) inherent in a DNA base-step (each atom has three spatial DOF) are reduced to six. 
Each base pair is represented by a plane; the six DOFs correspond to the three translational and three rotational transformations required to superimpose a base pair on its neighbor in the  3' direction. 
By calculating the required transformations for a wide range of DNA structures in a variety of contexts (e.g. protein--bound, in solution, etc.), one can construct a 6$\times$6 covariance matrix, $M$, for each unique base-step. 
This covariance matrix can be used to estimate the conformational volume of the base-step, hereafter referred to as ``$S$,'' as well as the force constants that penalize deformations of the six degrees of freedom for each base-step\cite{Olson1998, Olson2006}.
In short, this analysis can give extensive insight into the local flexibility of each base-step in a global sense through $S$ or in great detail through information about the flexibility of each degree of freedom.

In 3SPN.2C, a local orthogonal coordinate system or ``triad'' is defined for each base pair, as shown in Fig. \ref{confvol}.
The six DOFs mentioned previously are then the three translational and rotational transformations required to move from one base pair to the next base pair that constitutes a base-step. 
The triad for each base pair is defined as follows:
first, a vector is drawn from the base site of the sense strand to the base site of the antisense strand. 
This constitutes the ``$y$'' axis of the triad. 
Second, the ``$z$'' axis is constructed by drawing a vector from the center-of-mass of the two base sites that constitute the base pair that points in the 3' direction along the helical axis.
In practice this is done by finding the projection of the ``$y$'' axis on the helical axis; 
the ``$z$'' axis is then a unit vector parallel to the vector rejection of the ``$y$'' axis on the helical axis. 
Lastly, the ``$x$'' axis is the cross product of the previous two vectors ($\hat{\textbf{x}} = \hat{\textbf{y}} \times \hat{\textbf{z}}$) and points in the direction of the major groove. 
This is done for each base pair in a sequence and the six parameters for each base-step are determined using the method described by Calladine, making use of the so-called ``mid-step triad''\cite{ElHassan1995}. 
Using an ensemble of configurations from direct simulations, the covariance matrix is determined for each base-step. 
The conformational volume, $S$, of each base-step is calculated by taking the square root of the determinant of $M$. 
$S$ provides a measure of the flexibility of a base-step (more accurately, it provides an estimate of the variability in the six degrees of freedom for the base-step), with a larger value indicating greater flexibility. 
(Interested readers are referred to Ref. \onlinecite{Olson2006} for additional details).

\begin{figure}
%\begin{figure*}
  \begin{center}
    \includegraphics[width=8.50cm]{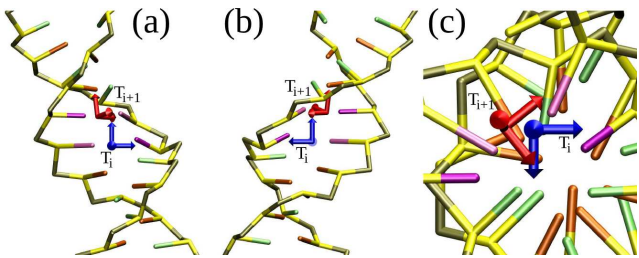}
    \caption{%
      Base-step triads for two base pairs comprising a base-step. T$_{\mbox{i}}$ and
	T$_{\mbox{i+1}}$ correspond to base pair triads at the 5' and 3' ends of a
	base-step, respectively. The conformational volume, $S$, of a base-step is
	determined from the covariance matrix of the six degrees of freedom of a
	base-step, as described by Olson {\it et al.}\cite{Olson2006}. The degrees of freedom are
	determined from the base-step triads according to the methodology of Calladine
	{\it et al.}\cite{ElHassan1995}. The T$_{\mbox{i}}$ triad is colored
	blue and the T$_{\mbox{i+1}}$ is colored red.
	%\textbf{Shouldn't the triad $T_{\text{i+1}}$ be on the bottom?}
    }
    \label{confvol}
  \end{center}
%\end{figure*}
\end{figure}

In order to incorporate the experimental values of $S$ in 3SPN.2C, we set the force constants of the Base--Sugar--Phosphate (B--S--P, where B=\{A, T, G, or C\}) and Sugar--Phosphate--Base (P--S--B) bending angles in the following manner:
\begin{equation}
  k_{\theta_{\tiny{i}}} = 120  + 340  \left( \frac{\sqrt{4.1} - \sqrt{S_{\mbox{\tiny{i}}}}}{\sqrt{4.1} - \sqrt{0.4}} \right)\mbox{kJ mol}^{-1}\mbox{ rad}^{-2}
  \label{eq:bends}
\end{equation}

\noindent where the index $i$ indicates a specific base-step, $S_{i}$ is the configurational volume of that step as characterized by Olson and coworkers\cite{Olson2006}, and 0.4 and 4.1 are the minimum and maximum values of $S_{i}$, normalized by $S_{\mbox{\tiny{AT}}}$, given by Olson {\it et al.} 
This limits values of $k_\theta$ between $120 \mbox{kJ mol}^{-1}\mbox{ rad}^{-2}$ and $460 \mbox{kJ mol}^{-1}\mbox{ rad}^{-2}$.
The choice of this function and the range of force constants are arbitrary but were found to give semi-quantitative agreement between simulation and experiment as shown in Fig. \ref{fig:validation}b.
Table \ref{bending_angles} in the Appendix provides the values of bending penalties used in the model.

\section{Results}

% Figure
\begin{figure*}
  \begin{center}
    \includegraphics[width=17cm]{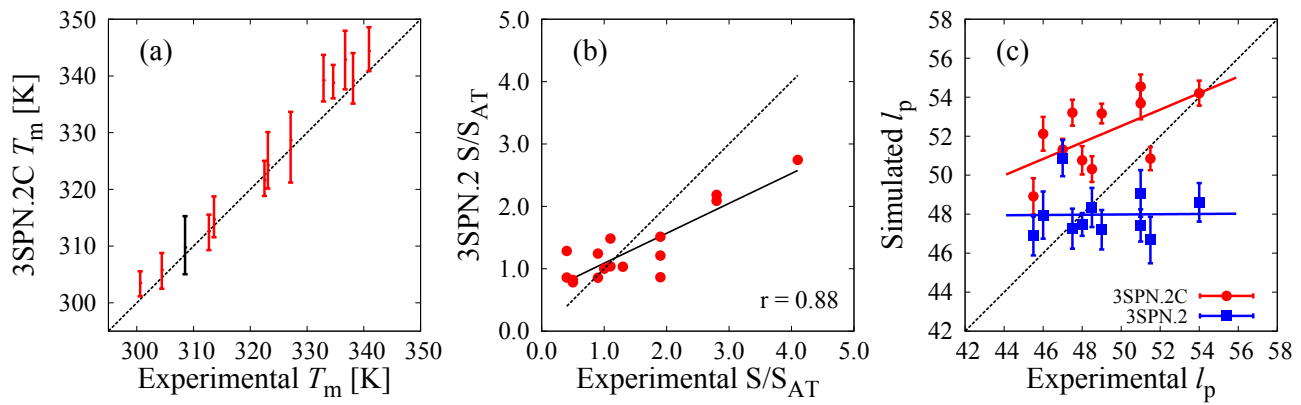}
    \caption{%
    (a) Consistency between melting temperatures $T_m$ calculated using the 3SPN.2C CG DNA model.
    The black error bar represents the predicted temperature of the sequence for which the model parameters were adjusted (5'-TACTAACATTAACTA-3'; $I$ = 69 mM).
    The red error bars represent the melting temperatures predicted for other sequences and ionic strengths.
    (b) Semi-quantitative agreement between experimental\cite{Olson2006} and simulated local flexibility $S$.
	To facilitate comparison of our data to that of Olson {\it et al.}, all values of $S$ in the figure are normalized by $S_{\mbox{\tiny{AT}}}$, as done in Ref. \onlinecite{Olson2006}. 
    The solid black line represents a best-fit to $S$.
    (c) Comparison of the ability of each CG model to capture the effect of sequence on collective flexibility via the persistence length $l_p$.
    Experimental persistence lengths are from Ref. \onlinecite{Geggier2010}.
    The solid lines represent linear fits to the data and provide a guide to the eye.
    In all figures, the dotted line represents exact correspondence between simulation and experiment.
    }
    \label{fig:validation}
  \end{center}
\end{figure*}

%% DUMMY Melting temperature

\subsection{Melting Temperature}\label{Sec:Tm}

While the 3SPN.2C model is intended to primarily simulate dsDNA, it is important to preserve the ability of dsDNA to melt.
Melting temperature calculations were performed using multiple walker metadynamics\cite{Raiteri2006}, as done previously\cite{Hinckley2013}.
Interstrand base pair interactions were scaled uniformly until good agreement was achieved between experimental and simulated melting temperatures for a reference sequence.
The melting temperatures of several other sequences and ionic strengths where then predicted.
Figure \ref{fig:validation}a demonstrates good agreement, as expected given the relatively minor modifications to the 3SPN.2 model.
This result also highlights that, for a single validation metric, there exist many sets of CG parameters that provide satisfactory performance.

\subsection{Sequence-dependent Flexibility}

\subsubsection{Local Flexibility}

To assess the behavior of the model with respect to local flexibility, $S$, we compare 3SPN.2C to the data of Olson and coworkers\cite{Olson2006}. 
%Langevin dynamics simulations were performed at 300 K and 150 mM in the NVT ensemble for each possible DNA tetramer.
Simulations were performed for each possible DNA tetramer and the local flexibility $S$ of each base step was calculated by averaging over the sixteen tetramers centered on the base step.
The agreement between 3SPN.2C and experimental values of $S$,  shown in Fig. \ref{fig:validation}b, is semi-quantitative, with a Pearson Correlation Coefficient of 0.88. 
Quantitative agreement is difficult to achieve because of the approximate methods used to define the base-step triads and the number of parameters that act in concert to affect local flexibility. 
It is possible to alter the parameters governing the anisotropy of the stacking and cross-stacking interactions in 3SPN.2C and so tune the local flexibility;
however, this requires extensive parameterization via multiple iterations in order to recover simultaneously stacking, melting, and bulk persistence length behaviors. 
We view the semi-quantitative level of agreement between model and experiment achieved here as satisfactory, given the difficulty in determining the six relevant degrees of freedom in the coarse-grained model.
%As this process can become prohibitively complex, we view the qualitative level of agreement between model and experiment achieved here as satisfactory.

%\begin{figure}
%  \begin{center}
%    \includegraphics[width=8.5cm]{./Fig3.eps}
%    \caption{%
%        Comparison of the local flexibility of 3SPN.2C ($y$-axis) with values from
%	Olson {\it et al.} ($x$-axis)\cite{Olson2006}. To facilitate comparison of our
%	data to that of Olson {\it et al.}, all values of $S$ in the figure are normalized
%	by $S_{\mbox{\tiny{AT}}}$, as done in Ref. \onlinecite{Olson2006}. Due to the
%	difficulties and assumptions inherent in determining $S$, we view this level of
%	agreement ($r = 0.88$) as reasonable.
%    }
%    \label{fig:flexres}
%  \end{center}
%\end{figure}
\subsubsection{Collective Flexibility}\label{Sec:Lp}

The persistence length of the 3SPN.2C representation of dsDNA is predominantly influenced by the strength of the bonded angle and dihedral potentials along the backbone.
The strengths of the angle potentials were assigned previously, with the other energy parameters as described in Ref. \onlinecite{Hinckley2013}.
The force constants of the Sugar--Phosphate--Sugar (S--P--S) and Phosphate--Sugar--Phosphate (P--S--P) angles along the DNA backbone, as well as the backbone dihedral force constants, were assigned using the experimental persistence length data of Geggier and Vologodskii\cite{Geggier2010}.
They determined the persistence length of $\approx$ 200 base pair DNA segments using cyclization assays and used the results to assign a bending penalty to each base-step. 
The S--P--S bending penalties in our model are given a force constant proportional to the base step bending penalties reported by Geggier and Vologodskii (Table \ref{bending_angles}). 
As with B--S--P and P--S--B bending penalties, the range of the force constant was chosen such that there was reasonable agreement with experimental persistence length data.
The force constants of the P--S--P bend and the dihedral potentials were modified until the persistence length of dsDNA came into good agreement with Ref. \onlinecite{Geggier2010}.

Persistence length calculations were performed using 75 base pair sequences taken from the middle of the sequences listed in Table S3 of Ref. \onlinecite{Geggier2010}.
Simulations were performed using both the 3SPN.2 and 3SPN.2C CG models.
The persistence length was calculated from the resulting trajectories using the helical axis autocorrelation function, as in previous work\cite{Hinckley2013}.
As demonstrated in Fig. \ref{fig:validation}c, 3SPN.2C is able to correctly capture trends in persistence length as a function of sequence.
In contrast, 3SPN.2 is not able to capture the trends in persistence length, despite the dependence of intrastrand stacking energies on sequence.
Such inability to capture trends in persistence length has also been observed in another coarse-grained DNA model with sequence-dependent base stacking energies\cite{Sulc2012}.

\subsection{Sequence-Dependent Minor Groove Width}

3SPN.2C simulations were performed in order to determine the minor groove width profiles of unbound DNA sequences.  
The minor groove width was determined from the resulting trajectories employing the method of El Hassan and Calladine\cite{ElHassan1995}.
In this method the minor groove width is defined by the distance between phosphate groups on opposing strands and refined to capture only the minimum separation between the two backbone curves.

To characterize the performance of 3SPN.2C with respect to sequence-dependent DNA shape, we employed three sources of experimental data. 
The first two rely on algorithms that predict the shape of the DNA minor groove based on experimental data employing photo-chemical cleavage of double-stranded DNA by the uranyl(IV) ion\cite{Lindemose2011} and hydroxyl radical cleavage (ORChID2)\cite{Bishop2011}, respectively. 
The third source of data is a library of DNA tetramers mined from the PDB databank by Rohs {\it et al.}\cite{Rohs2009}. 
In all cases, a comparison is made to the minor groove width determined through simulations of all 256 possible tetramers. 
Figure \ref{fig:orchid2} shows the correlation between 3SPN.2C and these three sources of experimental data. 
In panels (a) and (b), the middle 13 base-steps from tetrameric sequences are compared to model predictions ($N=3328$). 
In panel (c), only the minor groove width at the central base-step of the tetramer is compared to experimental data ($N=136$). 
In this third comparison, each experimental tetrameric minor groove width represents the average from all PDB structures containing the tetramer. 
These are the same data given in Ref. \onlinecite{Rohs2009}. 
In all cases, the correlation is statistically significant ($P < 10^{-12}$ in the worst case, the comparison to PDB data). 
However, the best correlation is between 3SPN.2C and the uranyl photo-cleavage model\cite{Lindemose2011}. 
%We also highlight minor groove widths that correspond to motifs centered on TA and GC dinucleotides, demonstrating the tendency of TA-centered tetramers to adopt narrower minor groove dimensions and GC-centered tetramers to adopt wider minor groove dimensions.

\begin{figure*}
  \begin{center}
    \includegraphics[width=17.0cm]{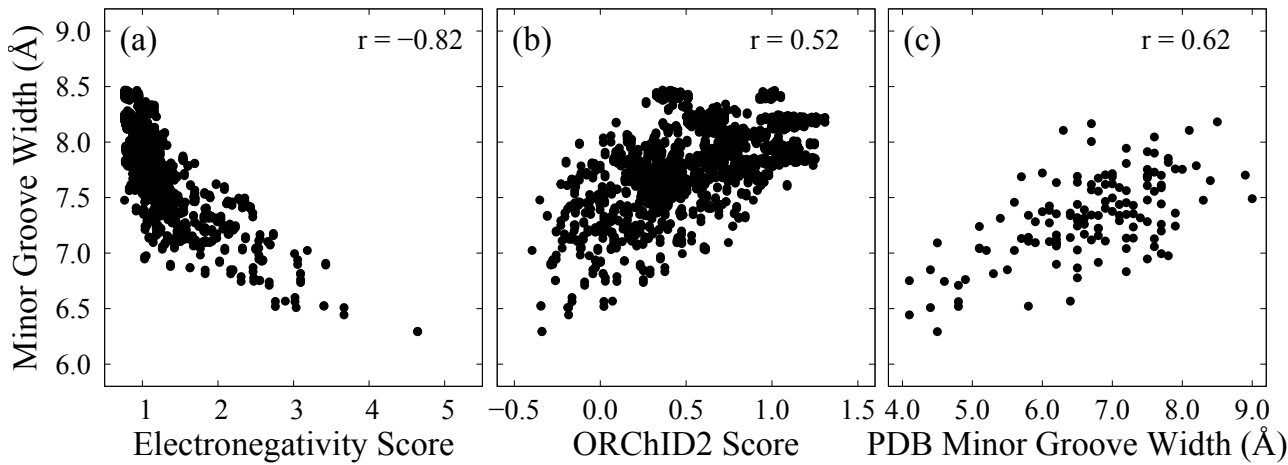}
    \caption{%
	Minor groove width from Langevin Dynamics simulations of all 256 possible tetramers compared to (a) Electronegativity score from uranyl(IV) ion\cite{Lindemose2011},
        (b) ORChID2 predictions\cite{Bishop2011} and (c) average minor groove widths from PDB structures\cite{Rohs2009};
    }
    \label{fig:orchid2}
  \end{center}
\end{figure*}

Representative examples of agreement between 3SPN.2C and the algorithms based on experimental cleavage data are given in Fig. \ref{fig:grooves}.
It is immediately clear that the qualitative agreement between 3SPN.2C and uranyl cleavage is better than that between 3SPN.2C and ORChID2.
This is in large part due to the noise in the ORChID2 signal. 
This is especially apparent in Fig. \ref{fig:grooves}e, in which both 3SPN.2C and uranyl cleavage predictions show an unchanging minor groove width for the sequence (CG)$_{10}$ while ORChID2 predicts an oscillating cleavage signal.

\begin{figure*}
  \begin{center}
    \includegraphics[width=17.0cm]{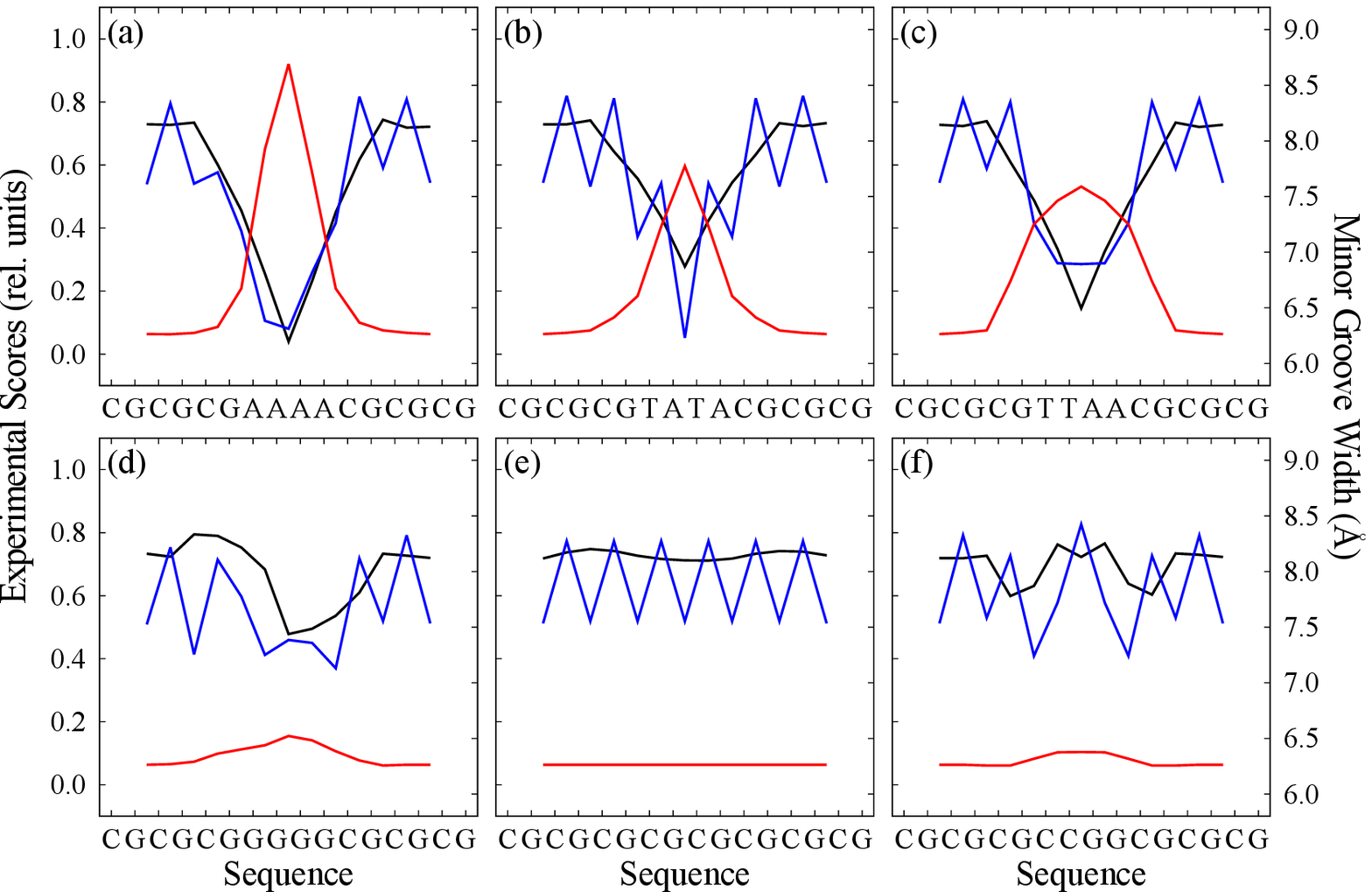}
    \caption{%
        Comparison of 3SPN.2C minor groove width profile (black lines) for six
	representative sequences to minor groove width prediction algorithms
	based on photo-chemical cleavage by uranyl (red lines)
	\cite{Lindemose2011} and hydroxyl radical cleavage (ORChID2, blue
	lines)\cite{Bishop2011}. Panels (a-c) show AA, AT, and TA motifs while
	panels (d-f) show CC, GC, and CG motifs.%
    }
    \label{fig:grooves}
  \end{center}
\end{figure*}

\section{Conclusion}

We have presented an extension of an existing coarse-grained DNA model for simulating DNA interactions with proteins.
The model uses experimental measurements of base-step parameters and mobility to inform the parameters imparting sequence-dependent shape and flexibility to dsDNA.
We have demonstrated that the model correctly predicts the effect of sequence on the persistence length of dsDNA.
We have also shown that the model is consistent with experimental measurements of minor groove width.

This model should find applications in scenarios where the structure and flexibility of DNA are important.
A forthcoming publication will demonstrate the successful use of the model to examine the origin of DNA affinity in the nucleosome.
The 3SPN.2C model has been implemented in LAMMPS\cite{Plimpton1995,LAMMPS} and is available online\cite{3SPN2}.

\begin{acknowledgments}
The University of Wisconsin--Madison Center for High Throughput Computing is gratefully acknowledged for providing computational resources and computer expertise.
We further acknowledge computational resources provide by the Midway computing cluster at the University of Chicago.
G.S.F. and J.P.L. gratefully acknowledge partial support of this research by the NSF-funded University of Wisconsin Nanoscale Science and Engineering Center (NSEC).
D.M.H. was funded by a Graduate Research Fellowship from the National Science Foundation (Grant No. DGE-1256259).
J.K.W. and J.J.D.P. acknowledge support from UChicago Argonne, LLC, Operator of Argonne National Laboratory (``Argonne''). 
Argonne, a U.S. Department of Energy Office of Science laboratory, is operated under Contract No. DE-AC02-06CH11357.  
%This work is supported by the National Science Foundation through the University of Wisconsin--Madison Nanoscale Science and Engineering Center \textbf{Grant No.?}.  
\end{acknowledgments}

\appendix*

\section{Shape-Dependent DNA Model Parameters}

Changing the topology of the 3SPN.2C requires modification to the equilibrium bond lengths, bend angles, and dihedral angles.
As stated in Section \ref{BaseStepParameters}, equilibrium DNA configurations are generated from X3DNA\cite{Lu2003} using parameters from Refs. \onlinecite{Morozov2009} and \onlinecite{Olson2006}. 
3SPN.2C is mapped onto the resulting atomistic structure according to the method outlined in Ref. \onlinecite{Hinckley2013}.
From this equilibrium structure, every bond, bend, and torsion in the structure is characterized and the relevant lengths and angles are written to files used as inputs for actual 3SPN.2C simulations. 
The code is available\cite{3SPN2} for generating 3SPN.2C topologies and performing simulations in LAMMPS using the modified energy parameters discussed below.

The energies of bonded and non-bonded interactions in 3SPN.2C differ from those in 3SPN.2. 
While the bond energies in 3SPN.2C are the same as 3SPN.2, the bending angles are different.
The sequence-dependent bending energies are shown in Table \ref{bending_angles}.
The magnitude of the dihedral energies are modified as follows:
$k_\phi$, the force constant for the Gaussian well-potential is modified from 6.0 kJ/mol/rad$^2$ to 7.0 kJ/mol/rad$^2$ and
$k_{\phi,\mbox{\tiny{periodic}}}$, the force constant of the additional dihedral potential (see Eq. \ref{eq:periodic}) is 2.0 kJ/mol/rad$^2$.

The 3SPN.2C model also includes changes to the base pairing, cross-stacking, and base stacking energies.
Base pairing and cross stacking interactions were scaled by a factor of 0.861.
The base pairing energies were 14.41 kJ/mol for A--T and 18.24 kJ/mol for G--C. 
The base stacking energies were calculated using metadynamics simulations (see Ref. \onlinecite{Hinckley2013} for additional details).
The energies of base stacking and cross stacking interactions are provided in Table \ref{base-energies}.
The other parameters that appear in the angle-dependent potentials ($K$, $\alpha$; see Ref. \onlinecite{Hinckley2013} for additional details) are unchanged from 3SPN.2.

\begin{table*}
  \small
  \caption{
        Bending angle energy constants employed in the 3SPN.2C DNA model.
        Single letter names for sites are as follows: A, T, C, and G denote the four DNA bases, S denotes a sugar moiety, and P denotes a phosphate moiety.
        B--S--P and P--S--B bend energies were assigned using Eq. \ref{eq:bends}; S--P--S bend energies were assigned using base-step bend energies from Ref. \onlinecite{Geggier2010}.
  }
  \centering
  \begin{tabular}{@{\vrule height 10.5pt depth4pt width0pt}lcc|lcc|lcc}
    \\[1ex]
    \hline
    Angle & Base-Step & $k_\theta$ & Angle & Base-Step & $k_\theta$ & Angle & Base-Step & $k_\theta$ \\
     & & kJ/mol/rad$^{2}$ & & & kJ/mol/rad$^{2}$ & & & kJ/mol/rad$^{2}$ \\
    \hline
    A-S-P & AA & 460 & C-S-P & CA & 206 & S-P-S & AA & 355 \\
    A-S-P & AT & 370 & C-S-P & CT & 358 & S-P-S & AT & 147 \\
    A-S-P & AC & 442 & C-S-P & CC & 278 & S-P-S & AC & 464 \\
    A-S-P & AG & 358 & C-S-P & CG & 278 & S-P-S & AG & 368 \\
    P-S-A & AA & 460 & P-S-C & AC & 442 & S-P-S & TA & 230 \\
    P-S-A & TA & 120 & P-S-C & TC & 383 & S-P-S & TT & 355 \\
    P-S-A & CA & 206 & P-S-C & CC & 278 & S-P-S & TC & 442 \\
    P-S-A & GA & 383 & P-S-C & GC & 336 & S-P-S & TG & 273 \\
    T-S-P & TA & 120 & G-S-P & GA & 383 & S-P-S & CA & 273 \\
    T-S-P & TT & 460 & G-S-P & GT & 442 & S-P-S & CT & 368 \\
    T-S-P & TC & 383 & G-S-P & GC & 336 & S-P-S & CC & 165 \\
    T-S-P & TG & 206 & G-S-P & GG & 278 & S-P-S & CG & 478 \\
    P-S-T & AT & 370 & P-S-G & AG & 358 & S-P-S & GA & 442 \\
    P-S-T & TT & 460 & P-S-G & TG & 206 & S-P-S & GT & 464 \\
    P-S-T & CT & 358 & P-S-G & CG & 278 & S-P-S & GC & 228 \\
    P-S-T & GT & 442 & P-S-G & GG & 278 & S-P-S & GG & 165 \\
    P-S-P & all & 300 & & & & & & \\
    \hline
  \end{tabular}
  \label{bending_angles}
\end{table*}

\begin{table}
\small
\caption{Base-stacking and cross-stacking energies for 3SPN.2C.  
Section (a) describes base-stacking energy scales.
Sections (b) and (c) describe cross-stacking energy scales.
Variables are as defined in Ref. \onlinecite{Hinckley2013}. 
 Upward-pointing arrows denote the sense strand while downward-pointing arrows denote the anti-sense strand (for cross-stacking interactions).}\label{base-energies}
\vspace{0.25 in}
\begin{tabular}{rc}
(a) &
\begin{tabular}{cc}
& \multicolumn{1}{c}{Base $^{3'}\uparrow$} \\
& $\epsilon$ \\ % & $\sigma$ & $\theta$ \\
& (kJ/mol) \\ %& (\AA) & ($^{\mbox{\tiny{o}}}$) \\
\text{Base} $_{5'}\uparrow$ &
\begin{tabular}{c|rrrr}
  & A & T & G & C \\ \hline
A & 13.82 & 15.05 & 13.32 & 15.82 \\
T &  9.15 & 12.44 &  9.58 & 13.11 \\
G & 13.76 & 14.59 & 14.77 & 15.17 \\
C &  9.25 & 12.42 &  8.83 & 14.01 \\
\end{tabular} 
%\end{tabular} &
%\begin{tabular}{c|rrrr}
%  & A & T & G & C \\ \hline
%A & 3.58 & 3.56 & 3.85 & 3.45 \\
%T & 4.15 & 3.93 & 4.32 & 3.87 \\
%G & 3.51 & 3.47 & 3.67 & 3.42 \\
%C & 4.15 & 3.99 & 4.34 & 3.84 \\
%\end{tabular} &
%\begin{tabular}{c|rrrr}
%  & A & T & G & C \\ \hline
%A & 100.13 &  90.48 & 104.39 &  93.23 \\
%T & 102.59 &  93.32 & 103.70 &  94.55 \\
%G &  95.45 &  87.63 & 106.36 &  83.12 \\
%C & 102.69 &  96.05 & 100.46 & 100.68 \\
%\end{tabular}
\end{tabular}\\ \\ \hline \\
(b) &
\begin{tabular}{cc}
& \multicolumn{1}{c}{Base $\downarrow^{5'}$} \\
& $\epsilon$ \\ %& $\sigma$ & $\theta$ \\
& (kJ/mol) \\%& (\AA) & ($^{\mbox{\tiny{o}}}$) \\
\text{Base} $_{5'}\uparrow$ &
\begin{tabular}{c|rrrr}
  & A & T & G & C \\ \hline
A & 1.882 & 2.388 & 2.439 & 1.680 \\
T & 2.388 & 1.882 & 2.187 & 2.566 \\
G & 2.439 & 2.187 & 3.250 & 0.972 \\
C & 1.680 & 2.566 & 0.972 & 4.135 \\
\end{tabular} 
%\end{tabular} &
%\begin{tabular}{c|rrrr}
%  & A & T & G & C \\ \hline
%A & 6.24 & 6.77 & 6.27 & 6.84 \\
%T & 6.77 & 7.21 & 6.53 & 7.08 \\
%G & 6.27 & 6.53 & 5.74 & 6.86 \\
%C & 6.84 & 7.08 & 6.86 & 6.79 \\
%\end{tabular} &
%\begin{tabular}{c|rrrr}
%  & A & T & G & C \\ \hline
%A & 154.04 & 158.77 & 153.88 & 157.69 \\
%T & 148.62 & 155.05 & 147.54 & 153.61 \\
%G & 153.91 & 155.72 & 151.84 & 157.80 \\
%C & 152.04 & 157.72 & 151.65 & 154.49 \\
%\end{tabular}
\end{tabular} \\ \\ \hline \\
(c) &
\begin{tabular}{cc}
& \multicolumn{1}{c}{\text{Base} $\uparrow^{3'}$} \\
& $\epsilon$ \\%& $\sigma$ & $\theta$ \\
& (kJ/mol) \\%& (\AA) & ($^{\mbox{\tiny{o}}}$) \\
Base $\downarrow_{3'}$ &
\begin{tabular}{c|cccc}
  & A & T & G & C \\ \hline
A & 1.882 & 2.388 & 2.566 & 2.187 \\
T & 2.388 & 1.882 & 1.680 & 2.439 \\
G & 2.566 & 1.680 & 4.135 & 0.972 \\
C & 2.187 & 2.439 & 0.972 & 3.250 \\
\end{tabular} 
%\end{tabular} &
%\begin{tabular}{c|cccc}
%  & A & T & G & C \\ \hline
%A & 5.58 & 6.14 & 5.63 & 6.18 \\
%T & 6.14 & 6.80 & 6.07 & 6.64 \\
%G & 5.63 & 6.07 & 5.87 & 5.66 \\
%C & 6.18 & 6.64 & 5.66 & 6.80 \\
%\end{tabular} &
%\begin{tabular}{c|cccc}
%  & A & T & G & C \\ \hline
%A & 116.34 & 119.61 & 115.19 & 120.92 \\
%T & 107.40 & 110.76 & 106.33 & 111.57 \\
%G & 121.61 & 124.92 & 120.52 & 124.88 \\
%C & 112.45 & 115.43 & 110.51 & 115.80 \\
%\end{tabular}
\end{tabular} \\ \\

\end{tabular}
\end{table}

\end{document}